\documentclass[
 amsmath,amssymb,
 aps,
floatfix
]{revtex4-2}

\usepackage{graphicx}
\usepackage{dcolumn}
\usepackage{bm}

\begin{document}

\preprint{APS/123-QED}

\title{Broadband Cavity-Enhanced Kerr Comb Spectroscopy on Chip}

\author{Andrei Diakonov}
\author{Konstantin Khrizman}%
\author{Eliran Zano}%
\author{Liron Stern}%
 \email{liron.stern@mail.huji.ac.il}
\affiliation{%
 Institute of Applied Physics, The Hebrew University of Jerusalem, Jerusalem, 91904, Israel
}%





\begin{abstract}
The broad and equidistant spectrum of frequency combs has had a profound impact on spectroscopic studies. Particularly, experiments involving the coupling of frequency combs to cavities have already enabled unprecedented broadband and sensitive spectroscopy on a single-molecule level. The emergence of integrated, compact and broadband Kerr-microcombs holds promise to bring many metrological and spectroscopic studies outside of the lab. However, performing cavity-enhanced direct frequency comb spectroscopy on-chip has remained a challenge. Here, we couple a microcomb source with a microcavity to extend the advantages of cavity-enhanced spectroscopy to photonically integrated circuits. By harnessing the coherent nature of the Kerr-comb and high-Q microcavity enhancement, we obtain a detailed dispersion landscape of the guided-wave mode and comprehensive frequency-dependent cavity lineshapes. Our microcomb-cavity coupling can facilitate photonically integrated cavity-enhanced biochemical spectroscopy by evanescently coupling analytes to the cavity's guided mode, a mode of operation we analyze and provide guidelines for. Demonstrated detailed dispersion measurements, overperforming state-of-the-art table-top tunable lasers in available bandwidth, show potential for integrated nonlinear optics applications, as precise dispersion management is crucial for such processes. Our chip-scale comb-cavity coupled platform suggests an integrated, broadband, cost-effective and accurate tool for the nonlinear optics studies as well as for ultra-compact bio- and chemical- sensing platform. 
\end{abstract}

\maketitle


\section{Introduction}

The direct interaction of laser light with different states of matter constitutes the cornerstone of modern spectroscopic studies. By confining matter within a cavity, a multitude of important spectroscopic modalities are made possible, such as enhanced sensing at the single-molecule level \cite{maccaferri2021recent}, dramatic amplification of optical nonlinearities \cite{bravo2007enhanced, lin2016cavity, fryett2017cavity} and the establishment of strong coupling for coherent energy flow between matter and the cavity \cite{walther2006cavity}. The advent of optical frequency combs (OFCs) provides a rich array of spectroscopic tools, presenting a particularly intriguing avenue for cavity-enhanced measurements. OFCs, consisting of ultra-wideband coherent spectra in the form of equidistant sharp peaks, have become an essential metrological tool to bridge the gap between optical and RF frequencies \cite{cundiff2003colloquium}, thus unveiling important applications in timekeeping \cite{ludlow2008sr} and frequency synthesis \cite{spencer2018optical}. The expansion of the utility of OFCs beyond coherent up and down conversions has allowed them to play a significant role in a multitude of novel applications such as spectroscopy \cite{foltynowicz2011optical}, dimensional metrology \cite{coddington2009rapid, schuhler2006frequency} and environmental monitoring \cite{coburn2018regional}. Naturally, OFCs may perfectly complement cavities leading to the emergence of Cavity-Enhanced Direct Frequency Comb Spectroscopy \cite{thorpe2008cavityRev} (CE-DFCS), which allowed to probe complex molecular structures with an extremely broad bandwidth and high sensitivity \cite{spaun2016continuous}, investigate the kinetics of different radicals \cite{bjork2016direct, fleisher2014mid} and to provide a high-resolution spectroscopy of gases \cite{picque2019frequency} with a prominent application to human breath analysis \cite{thorpe2008cavity}. A further major development that broadens the capabilities of comb spectroscopy is the use of two combs with offset repetition rates, known as dual-comb spectroscopy \cite{coddington2016dual}. Essentially, this multi-heterodyne technique transfers the detection to the radio-frequency range, allowing for increased acquisition speed and eliminating the need for an optical spectrum analyzer \cite {bernhardt2010cavity,fleisher2016coherent}.

Recent years have shown the emergence of the field of integrated photonics, introducing the manipulation of light on a chip by utilizing various photonic components, such as waveguides, modulators, and detectors. This approach promises to revolutionize sensing, data communication and quantum technologies by enabling high-speed, low-power and compact photonic devices. Central to these technologies are integrated micro-cavities, enabling unprecedented levels of field confinement and enhancement, which in turn lead to novel applications in both sensing and non-linear optics. Various integrated cavity configurations have been implemented, such as  Bragg-mirror based Fabry-Perot cavities  \cite{pruessner2007integrated}, whispering gallery mode (WGM) resonators \cite{soltani2007ultra}, microring resonators (MRRs) \cite{bogaerts2012silicon}, and even the combination of thereof \cite{wu2017micro}. Arguably, one of the most intensively used and investigated platforms to date is SiN MRR configuration \cite{sun2022silicon} , which benefits from extremely low-loss, wide operational bandwidth and CMOS-compatible manufacturing process. Indeed, a plethora of sensing modalities have been demonstrated using integrated silicon-based micro-resonators, as for example concentration and temperature measurement of glucose solution \cite{kwon2008microring}, iron corrosion sensing \cite{ahmed2016optical}, and CES with microfluidic channels \cite{nitkowski2008cavity}, to name a few. Owing to to the inherent field enhancement in optical cavities, the threshold for nonlinearities reduces significantly, allowing to exploit $\chi^{(3)}$ process within the waveguide. In particular, Kerr-comb generation may be initiated by a cascaded four-wave mixing process producing coherent and octave-spanning spectrum \cite{pfeiffer2017octave}. Consequently, microcombs have already been demonstrated to be extremely versatile in their applications including ranging \cite{suh2018soliton, lukashchuk2023chaotic}, sensing and dual-comb spectroscopy \cite{bao2021architecture}, optical clocks \cite{newman2019architecture} and precision atomic spectroscopy \cite{stern2020direct}.  The main challenge for reaching wideband comb generation and a fully-coherent soliton state is dispersion engineering \cite{fujii2020dispersion}. In order to compensate for material non-linearity through modal dispersion, the geometry of the MRR waveguide must be precisely adjusted. Simultaneously, the modal dispersion landscape defines the existence of dispersive waves, required to cover the octave range. This dictates the necessity for the accurate and broadband dispersion measurement of the cavity. 

Here, we demonstrate the coupling of  a photonically-integrated microcomb source with a microring resonator to showcase chip-scale cavity-enhanced spectroscopy.  By doing so, we introduce a universal  platform allowing  complete microcomb dispersion characterization and microcomb-cavity enhanced spectroscopy on-chip. In this study we use two identical MRRs which constitute the comb and the cavity. Simple thermo-optic tuning of the cavity, probed by a Kerr microcomb operating in a soliton regime, enabled us to thoroughly characterize its guided-mode dispersion by providing detailed and broadband cavity lineshapes and resonance frequency positions. Our results agree well with comprehensive numerical simulations as well as reference measurements. In our current demonstration, we were already able to provide a broad bandwidth of nearly 250 nm. The effective bandwidth measurement obtained is already highly competitive relative to contemporary table-top tunable laser sources such as ECDLs and OPOs. Ultimately, our approach can support even larger bandwidths approaching an octave span. Our demonstration exploits the coherence and the high-Q of the cavity to allow cavity enhanced spectroscopy of the guided-wave mode dispersion. By evanescently coupling molecules to the guided mode, we can extend the spectroscopic approach to allow cavity-enhanced direct frequency comb spectroscopy (CEDFCS) of anlaytes.  In particular, we numerically investigate the coupling of a microcomb to a water-cladded cavity, enabling an extremely broad-band, sensitive, and photonic-integrated platform for sensing. Our microcomb-cavity coupled system address key challenges in integrated photonics, enabling rapid and precise mapping of the microresonator dispersion landscape, while concurrently performing broadband resonantly enhanced spectroscopy. As such, it paves the way for a variety of comb-cavity experiments and applications in the fields of nonlinear integrated photonics, as well as environmental and biomedical sensing.

    \section{Comb-cavity coupling concept}
 
\begin{figure*}[!t]
            \includegraphics[scale=0.95]{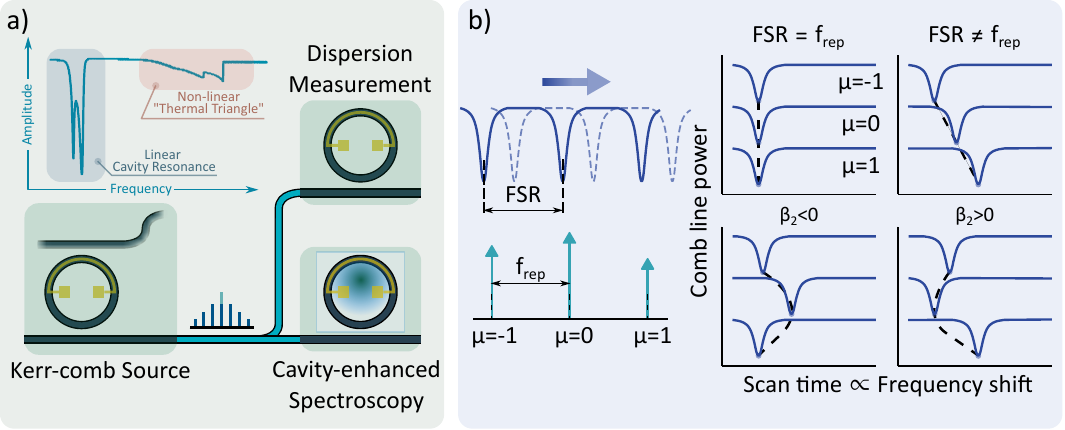}
            \caption{\textbf{Photonically-integrated comb-cavity coupling concept} a) Source Kerr-comb is coupled to either a silicon-dioxide cladded cavity, allowing waveguide-mode dispersion measurement or to an analyte-cladded cavity allowing for the evanescent-mediated specimen-light interaction. Heaters are used to set the appropriate pump wavelength of the comb source and to implement the temperature sweep of the target cavity. b) Schematic description of the mechanism: cavity resonances with a certain FSR shift in response to the temperature change, while the stationary comb with a linespacing of $f_{rep}$ acts as a probe signal. Change in the intensity of a specific comb line, resulting from the coupling of comb-tooth with a cavity resonance, provide a complete dispersion data - relative frequency positions of the resonances along with the corresponding lineshapes. Straight horizontal alignment of the resonances corresponds to a perfect dispersionless cavity with FSR $= f_{rep}$, linear one reveals the mismatch between FSR and $f_{rep}$, and parabolic arrangement indicates a second-order dispersion characterizing group velocity dispersion.}
            \label{fig:concept}
        \end{figure*}

In this part, we present the concept of photonically-integrated coupling of a microcomb and a microring resonator. As will be shown and discussed later, this concept can be simultaneously used for the direct cavity-enhanced frequency comb evanescent spectroscopy as well as provide an ultra-wideband measurement platform for thorough microcavity dispersion characterization. Our conceptual arrangement is schematically depicted in the Fig.\ref{fig:concept}(a). Operationally, a Kerr-comb spectrum is initiated in the source MRR and directly coupled to a target MRR. The latter frequency landscape is controlled by means of thermo-optic effect, which can be routinely controlled by integrated microheaters. Here, such an effect primarily alters the refractive index of the constituting MRR materials, thus, shifting the resonance frequencies. The Kerr-comb acts as a reference frequency source which tracks the moving resonance transmission dips of the target cavity. As shown in Fig.\ref{fig:concept}(b) cavity transmission response can be extracted from monitoring the power of a specific comb line, providing a detailed lineshape data and consequently revealing the resonance frequency  position, line-splitting, contrast and linewidth. Cavity transmission acts as a filter which is shifting in response to the temperature change effectively implementing the convolution operation with respect to the comb signal. Resulting data (Fig.\ref{fig:concept}(c)) naturally represents the so-called integrated dispersion $D_{int}$ up to a linear term, which results from a mismatch between the comb repetition rate and cavity free spectral range (FSR). We can define $D_{int}$ in terms of the resonance frequency positions with an angular number m $\omega_m = \omega_0 + D_1 \mu + \frac{1}{2}D_2 \mu^2 + ...$ as follows:
\begin{equation}
    D_{int} = \omega_m - \omega_0 - D_1\mu
    \label{eq:d_int}
\end{equation}
, where $D_1/2\pi$ is an FSR of the cavity in Hz, the rest of the terms define higher-order dispersion and $\mu$ is a longitudinal mode number with respect to the reference mode $\mu=0$. Our results are acquired through a continuous thermo-optical scan of the target cavity. Prior this process, we synchronize the resonant frequency of the target cavity with the microcomb's pump frequency by applying a fixed thermal offset. To determine the magnitude of this temperature offset, we measure the relative frequency position of the pumped source MRR resonance with respect to the resonance of the target cavity. This can be achieved by a laser frequency scan around the pump resonance. Knowing the frequency scan rate and the temperature-to-frequency resonance shift coefficient of the material, one can calculate the initial temperature value and subsequent temperature change range for the target cavity scan. Once calculated, the desired temperature profile is applied to the target cavity, and the corresponding frequency sweep range is then deduced. Above a certain power threshold, the resonant lineshape appears as a triangular-like (often coined 'thermal triangle') resonance instead of a Lorentzian one. This is  a consequence of a thermo-optic effect supported by the small mode volume and field enhancement, \cite{Carmon:04}. Thus, the photodetected signal of the scanning laser would consist of the source MRR thermal triangle and a conventional Lorentzian resonance as is shown in Fig.\ref{fig:concept}(a). Simultaneous presence of the thermal triangle along with the Lorentzian resonance neatly demonstrates the conceptual coexistence of the non-linear and linear effects in a comb-cavity coupled system.

Repetition rate $f_{rep}$ of the frequency comb defines the even spacing of the comb lines in the frequency domain, thus, scanning the target cavity resonances with respect to the stable comb lines results in the partial subtraction of the linear term defined in \ref{eq:d_int}. In case of a closely matched comb $f_{rep}$ and cavity FSR the remaining linear term can be negligibly small which would naturally result in an integrated dispersion data representation very close to its original definition.        
For the ideal scenario of a dispersionless cavity with an FSR precisely matching $f_{rep}$ the measurement would show the resonance positions perfectly aligned with respect to each other, which implies that the overlap between the comb lines and corresponding resonance dips during the shift occurred simultaneously. In case of a mismatch between a dispersionless FSR and identical $f_{rep}$, linear term would show up. Second-order dispersion, characterizing the group velocity dispersion (GVD) for the pulse propagation, would have a parabolic shape with a concavity defined by the normal or anomalous dispersion character. Analogously, if the higher-order dispersion is presented, it would be captured by the measurement. 
An access to the resonance lineshape data allows to use the aforementioned setup for implementation of a photonically integrated cavity-enhanced spectroscopy. Because such a method provides simultaneous information about $D_{int}$ (real part of the refractive index) as well as the losses (imaginary part of the refractive index and the resonance contrast) we term it as complex $D_{int}$ throughout this paper. By exposing the MRR's top cladding one can evanescently couple a portion of  the circulating light to a material under investigation for the purpose of cavity-enhanced sensing. For example, liquid or gas placed in the cladding of the MRR waveguide would alter the effective refractive index of the transversal mode propagating along the cavity and simultaneously induce frequency-dependent losses. By comparing the reference complex $D_{int}$ with the one obtained after adding the material under the investigation to the sensing cavity, absorption and dispersion spectroscopy can be implemented. Integrated comb-cavity coupling constitutes a conceptually universal, broadband and compact platform for cavity-enhanced direct frequency comb spectroscopy and microcavity studies. 
    
    \section{Experimental results}
\begin{figure*}[h!]
                    \includegraphics[scale = 0.75]{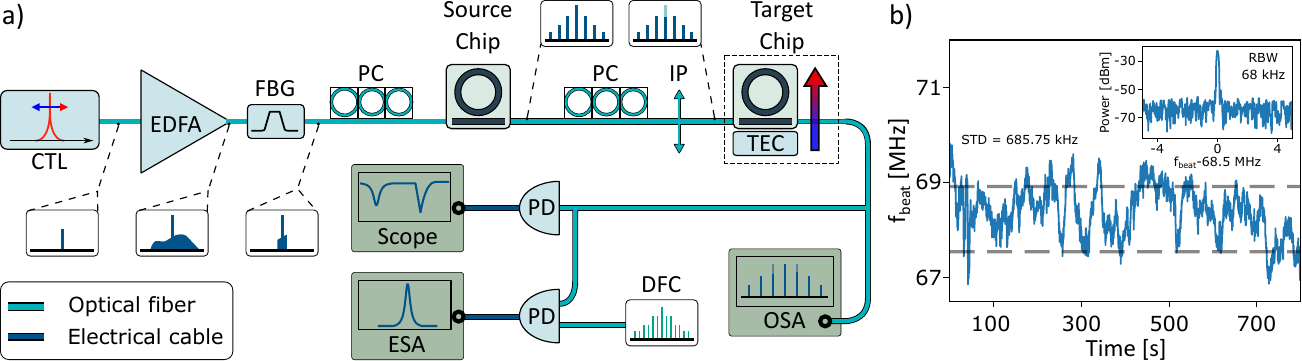}
                    \caption{\textbf{Chip-to-chip comb-cavity coupling experimental realization}  a) Continuously tunable laser (CTL) is amplified by an Erbium-Doped Fiber Amplifier (EDFA) with an amplified spontaneous emission filtered by a band-pass Fiber Bragg Grating (FBG) filter. The first polarization controller (PC) is used then to optimize the pump coupling to the SiN chip for the soliton generation achieved by a forward-backward tuning of the CTL in the vicinity of the source MRR resonance. The second PC in conjunction with an in-line polarizer (IP) is used to eliminate the pump to prevent triggering non-linear effects in the target cavity. Our target chip is placed on a copper stage with the attached Peltier element for the temperature sweep, which is controlled by means of Thermo-Electric Controller (TEC). Resulted comb signal is split for the detection of the individual comb line power by an Optical Spectrum Analyzer (OSA) simultaneously with monitoring the chip's temperature. Another part of the signal is photodetected for facilitating the soliton generation, and the last portion is used for detecting the beat between the coherent comb state and table-top Difference Frequency Comb (DFC). b) Comb coherence verification by measuring the beat frequency between the comb line and DFC. Standard deviation of 700 kHz is obtained for the 800 s measurement with a sub-MHz  linewidth.}
                    \label{fig:setup}
                \end{figure*}

    The schematic representation of our experimental setup is illustrated in Fig.\ref{fig:setup}. To implement comb-cavity coupling we used two SiN chips with identical design parameters. Yet, owning to manufacturing tolerances, the optical path differs slightly and consequently the dispersion of the MRRs. The cross-sectional dimensions have a crucial role in defining the dispersion of the ring. In our work we targeted an anomalous dispersion that we achieved with the microring waveguide of 1.55 $\mu$m width  and 0.8 $\mu$m height for MRR and 1 $\mu$m width and 0.8 $\mu$m height for bus waveguide, correspondingly. A MRR radius of 22.5 $\mu$m resulted in approximately a 1 THz separation between the comb lines. 

    Precise comb-assisted spectroscopy requires the source to be coherent. Otherwise, the broad linewidth of comb-teeth \cite{del2011octave} as well as their corresponding jitter can hinder the measurement resolution and degrade the available signal-to-noise ratio. This requirement can be matched by initiating a soliton state. To that end, we employed a technique highly reminiscent of a backward tuning reported by Guo et.al. \cite{guo2017universal}. Amplified continuously tunable laser (CTL) was used to generate a multi-soliton state at first, which then was converted to the single-soliton state by the backward tuning. In our case no fast forward-tuning was required in order to get a multi-soliton state, which we attribute to the presence of  avoided mode-crossings around the pump resonance \cite{herr2014mode}. Fiber Bragg grating (FBG) based band-pass filter was used to filter out the amplified spontaneous emission of the Erbium-doped fiber amplifier (EDFA). The first polarization controller (PC) was used to achieve an optimal coupling of the pump to the chip, while the second PC in combination with an in-line polarizer allowed us to reduce the pump power to avoid triggering non-linear effects in the target MRR cavity. 
    
    To choose an appropriate temperature scan range and to control the laser offset during the soliton generation, optical signal was photodetected and monitored by an oscilloscope. We scanned the laser wavelength to obtain the transmission signal of the target cavity simultaneously with the thermal triangle of the source MRR to identify the initial value of the temperature sweep. In order to verify the coherence of the microcomb we heterodyned an adjacent line to the pump frequency with a table-top stabilized frequency-comb line. The resulting beat-note  was captured by an electrical spectrum analyzer (ESA). Sub-MHz linewidth of the beat verifies the coherent nature of the generated comb state. Fluctuation of the beat frequency during an 800 seconds measurement did not exceed a peak-to-peak level of 3 MHz with a standard deviation of approximately 700 kHz.
    As described in the previous section, after propagating a comb light through the MRR cavity we initiated the temperature sweep of the target chip. Target cavity chip was placed on a copper stage connected to a Peltier element, which in turn was managed by a thermo-electrical cooler (TEC) controller. An Optical Spectrum Analyzer (OSA) was used to monitor comb lines power, while simultaneous temperature tracking of the target chip was obtained using a thermistor. We used an apriori temperature-to-frequency shift calibration to convert the temperature measurement data into frequency shift values. 
                   
                \begin{figure*}[htp]
                \centering
                \hspace{-1 cm}
                    \includegraphics[scale=0.95]{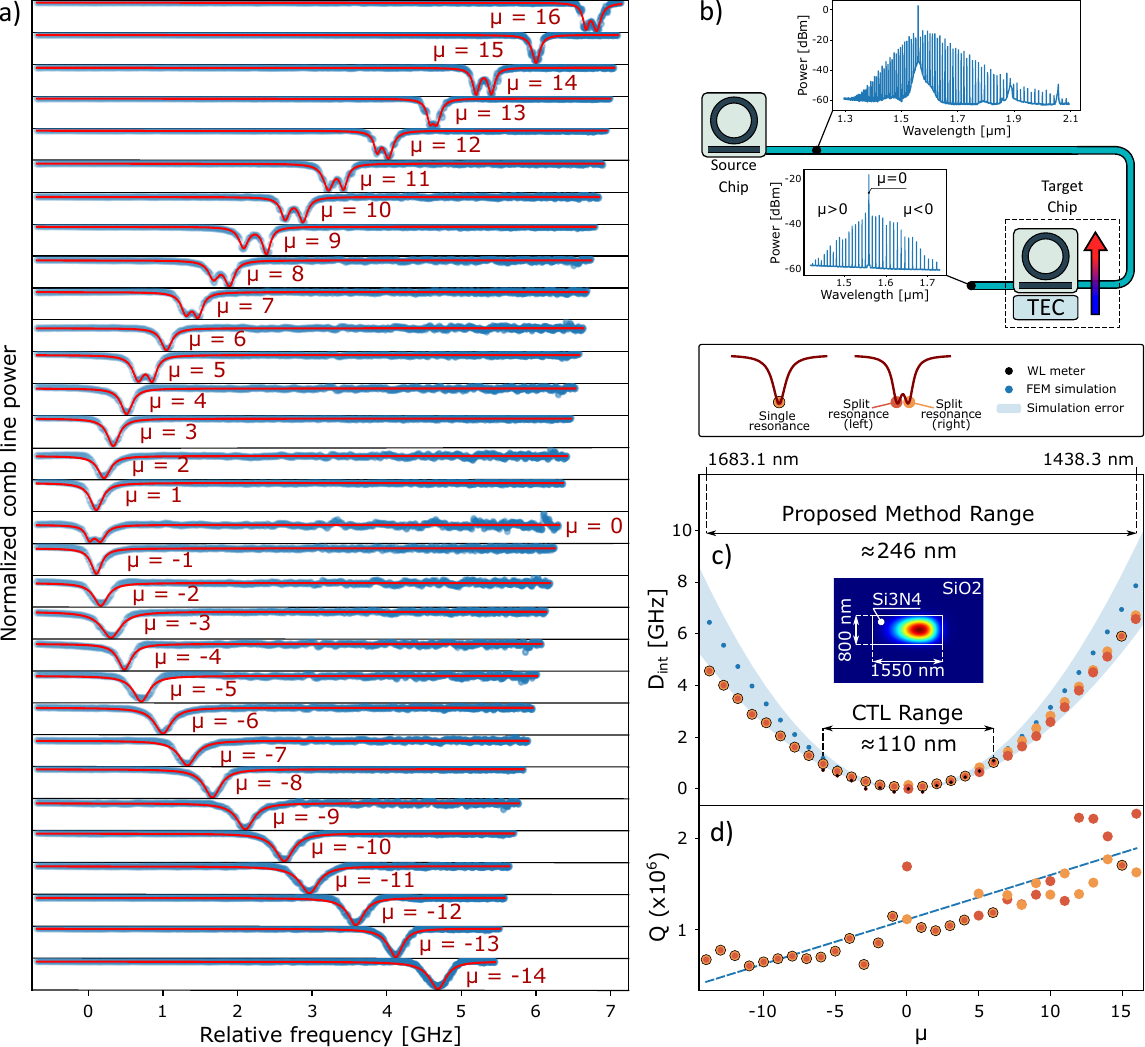}
                    \caption{\textbf{Microcomb enabled complex integrated dispersion of the high-Q MRR cavity} (a) Complex integrated dispersion measurement obtained from micocomb-cavity coupling and subsequent temperature scan. Each row corresponds to a specific comb line mode number $\mu$,  ordered with respect to the pump  (inset in (b) shows $\mu$ ordering with respect to the comb lines), and represents the cavity transmission as function of frequency. Light blue dots represent the raw measurement data with the superimposed (red curves) fitted Lorentzian lineshapes. (b) Reduced schematic of the setup where initial wide comb is shown in comparison with the comb state after the propagation through the target cavity. (c) Real integrated dispersion, extracted from the complex data (a). The peaks of the split resonance are shown in light and dark orange, and a single resonance is designated by the combination of thereof. Blue dots show the result of a FEM simulation for $D_{int}$ with the blue light background representing uncertainty in the WG dimensions. Inset demonstrates the  WG corss-sectional dimensions along with the basic TE mode intensity acounting on the waveguide curvature. Wavelength meter measurement is depicted by black dots and the obtained CTL range is shown (110 nm). (d) Q-factor distribution extracted from (a) demonstrating the coupling-loss dependent loaded Q-factor which is a consequence of our point coupling between the bus waveguide and MRR. Measurement range obtained in the current study exceeds state-of-the-art CTL range more than twice as much (246 nm).}
                    \label{fig:results}
                \end{figure*}
                
Figure \ref{fig:results} shows the obtained complex integrated dispersion measurement and its processing. Our single-shot measurement has the capacity to resolve 31 cavity lines corresponding to a frequency (wavelength) deviation of 31 THz (246 nm).  Each row in Fig.\ref{fig:results}(a) is assigned a mode number $\mu$ with respect to the comb pump mode number $\mu_p=0$, where $\mu$ decreases with the increasing wavelength and increases with respect to the frequency.  For clarity we show the mode numbers related directly to the comb structure used in the experiment in Fig.\ref{fig:results}(b). Raw measurements from the temperature scan are shown with light blue dots with the processed fit data (red curve) on top of that. Apparent slope in the demonstrated data of Fig.\ref{fig:results}(a) is due to the frequency-temperature coefficient (GHz/$^{\circ}$C) frequency-dependent nature, the measurement was calibrated with respect to it. The initial data in Fig.\ref{fig:results}(a) was acquired simultaneously, after which every row was multiplied by a corresponding frequency dependent temperature-to-frequency conversion coefficient resulting in the scaling of the y-axis and appearance of the aforementioned slope.
From the complex $D_{int}$ picture we extracted the ordinary $D_{int}$, which characterizes cavity modes shift with respect to their dispersionless evenly spaced positions. The dominant parabolic nature of the frequency shifts shown in Fig.\ref{fig:results}(b) and their concavity indicate the anomalous dispersion. The second-order fit integrated dispersion coefficient $D_2/2\pi$ was calculated to be 24.8 MHz. To back-up our results we separately measured  $D_{int}$ using a wavelength meter, the result is depicted by black dots in Fig.\ref{fig:results}(b). It is instructive to notice that the available measurement range obtained with a state-of-the-art CTL and a wavelength meter (110 nm) is more than twice as small as the range provided by the temperature scan (246 nm). We also emphasize that our setup (Fig.\ref{fig:setup}) was not optimized in terms of the available comb probe signal bandwidth as we will discuss below. To further verify the validity of the proposed method we executed numerical simulation by using a finite-elements-method (FEM) mode solver. The refractive index from the simulation was used to calculate $D$ coefficients for the integrated dispersion \cite{twayana2022comb}. The simulated $D_{int}$ (blue dots) is in a good agreement with the measurements obtained, and the light blue background represents the uncertainty range caused by  manufacturing imperfections ($\pm$ 20 nm variation in the height and width of the waveguide). The second-order dispersion coefficient $D_2/2\pi$ obtained from the simulation was equal to 31.2 MHz, which is in a good agreement with the measurement results, given the uncertainties in refractive indices values, and geometry. 
Apart from the frequency positions of the resonances, we are also able to to extract the quality factors of the corresponding cavity resonances, which is presented in Fig.\ref{fig:results}(d). The resulting data demonstrates a low-loss nature of the SiN waveguide based cavity with  Q-factors ranging approximately from 1 up to 2 millions. As a consequence of the the relatively high Q-factors, splitting of some lineshapes is evident (owning to defects), which we include individually in the plot. Another interesting feature of the Q-factor distribution is its increasing trend with the increase of the mode number. We attribute this behavior to the influence of the frequency-dependent point coupling between the bus waveguide and the MRR itself, which is consistent with the study of Moille et.al. \cite{moille2019broadband}. 
 
Complex integrated dispersion representation is naturally appealing for the cavity-enhanced spectroscopy applications. For the case of the integrated photonic chip platform the sensing cavity can be realized by exposing the cladding of the target MRR. The evanescently coupled guided wave would be perturbed in the presence of a specimen such as a gas or liquid, resulting in a modified frequency-dependent effective refractive index. Comparing the reference complex $D_{int}$ obtained without the specimen with $D_{int}$ acquired after adding the specimen enables the identification of the spectral components of the material under investigation. 

High-Q cavities used in our experiment provide a potential platform for dramatic sensitivity enhancement for spectroscopy experiments. Signal-to-noise ratio improvement due to the presence of the low-loss cavity should be approximately proportional to its Finesse, which defines the number of roundtrips the light field would make until its energy would decay to the $1/e$ with respect to the initial value. To verify this concept in context of our proposed platform we performed rigours calculations where we used water as the cladding of our cavity.

        \begin{figure*}[b]
        \hspace{-0.5cm}
            \includegraphics[scale=0.95]{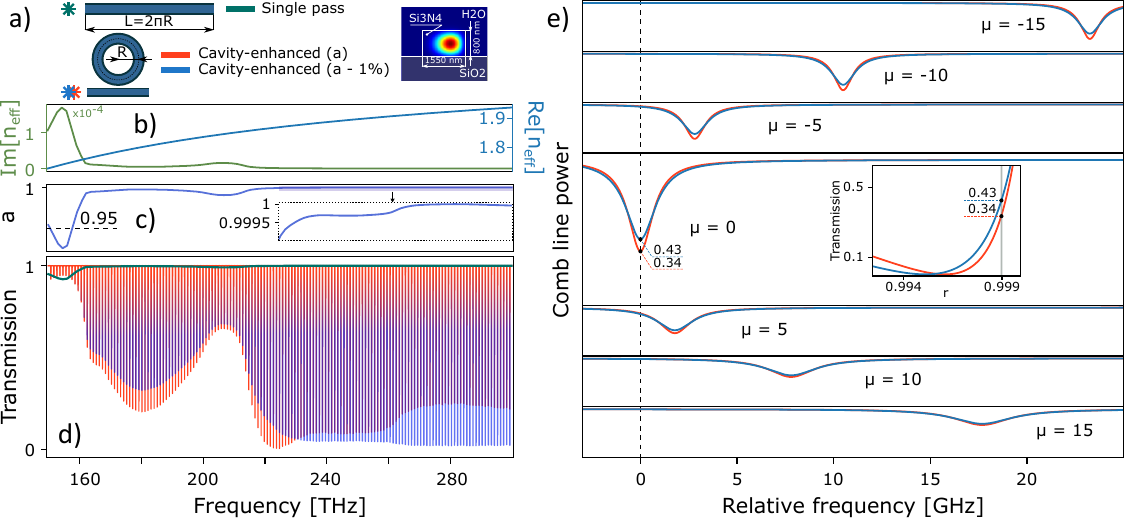}
            \caption{\textbf{Cavity-enhanced comb-cavity water spectroscopy}: (a)Schematic comparison between a single-pass waveguide approach and a cavity-enhanced approach with a sensing MRR. Here, the waveguide length is $L = 2\pi R$, where $R$ is the MRR radius. In both cases the waveguide upper cladding is exposed, which allows the specimen to be evanescently coupled to the guided mode.  (b) Real and imaginary part of the $n_{eff}$  obtained from the FEM simulation for the waveguide with water cladding. (c) Loss coefficient of the water with the zoom-in view which demonstrates subtle changes on a smaller scale. (d) Comparison between the sensitivity of the single-pass approach with respect to MRR enhancing cavity. Only substantial material losses (close to 10$\%$ ) can be resolved with a single-pass approach with reasonable contrast (i.e. > few percent) (dark blue curve), while cavity-enhancement (orange) reveals even the minute changes on the scale less than 0.1$\%$. Blue curve represents the change of 1$\%$ in water losses which mimics the scenario for biomedical sensing in liquid medium. (e) Complex $D_{int}$ extracted from the simulation, where $\mu = 0$  corresponds to 1559 nm (same as in our experiment). Much higher concavity is expected with respect to the data obtained from the in Fig, 3. Inset shows the dependence of the transmission on the coupling coefficient, for two cases of loss in the MRRs cladding, which defines the resulting signal (d) contrast.}
            \label{fig:cavity-enhancement}
        \end{figure*}
        
MRR cavity transmission can be mathematically represented by the following equation\cite{van2016optical}:
\begin{equation}
    H(\omega) = |(r - ae^{i\phi})/(1 - rae^{i\phi})|^2,
    \label{eq:transm}
\end{equation}
where $r$ is the self-coupling coefficient, $a$ designates cavity losses, $\phi = n_{eff}(\omega)kL$ is the phase of the waveguide mode with the wavenumber $k$ and effective refractive index $n_{eff}$ propagating along the cavity of length $L$. We calculated the $n_{eff}$ for the waveguide dimensions used in the experiment, but this time the SiO2 cladding was substituted with the water as is depicted in Fig.\ref{fig:cavity-enhancement}(a). Complex $n_{eff} = n_r + i n_i$ obtained from the simulation is shown in Fig.\ref{fig:cavity-enhancement}(b). From the imaginary part of $n_{eff}$ we calculated a loss coefficient as $a = \exp(2\pi n_i L f/c)$ that is shown in Fig.\ref{fig:cavity-enhancement}(b), the inset demonstrates a zoom-in view to emphasize the features presented at a small scale. Cavity transmission calculated with the Eq.\ref{eq:transm} using the losses coefficient obtained from the numerical calculation is shown in Fig.\ref{fig:cavity-enhancement}(c): orange plot demonstrates the transmission with the initial water $n_{eff}$ data, while the blue one shows the effect of the $0.1\%$ change in the absorption coefficient $a$. Induced absorption change in the simulation mimics the scenario when liquid is used as a mediator between the biological measurands like for example biomolecules, DNA or RNA and light. This became especially important in the context of the recent COVID-19 pandemic to develop a reliable and compact sensor for the virus signature detection \cite{grosman2023chip}. We want to emphasize that for the presented simulation, we used a fixed value of $r=0.993$, which, due to the frequency-dependent nature of the loss coefficient, resulted in critical coupling (zero transmission value) only for a specific portion of the transmission spectrum.  This suggests that bus-ring coupling can be used as an additional degree of freedom during the design of a specific sensing device. To show the sensitivity enhancement of the absorption spectroscopy we also plot the transmission spectrum of the straight waveguide with length L equal to the MRR's circumference as a conventional spectroscopy approach reference. It is clearly seen that the use of cavity allows to distinguish minute features of the water absorption spectrum, while a single-pass approach is sensitive only to a huge variations of losses. We also note that the information about the change in the real part of the refractive index is merely absent for the case of the waveguide sensing, such a setup should be supplemented with a form of interferometric subsystem \cite{dulkeith2006group, dwivedi2015measurements} in order to get an access to it. Quantitatively, we define cavity-enhancement as the ratio between absorption per unit length in the straight waveguide with respect to the effective absorption per unit length in MRR, which is a frequency dependent quantity. Based on the simulation we evaluated the effective enhancement to be 39 at 150.7 THz (1.99 $\mu$m) and 9200 at 230.6 THz (1.3 $\mu$m) for the two typical cases of a high and small losses. These results suggest that in case of a small change in the losses the effective enhancement can reach nearly 4 orders of magnitude, exploiting the high finesse of our cavity. Fig.\ref{fig:cavity-enhancement}(d) shows the complex $D_{int}$ obtained from the simulation for the same mode numbers $\mu$ that were used in the experiment. In the presence of water, the concavity of the second-order dispersion is significantly increased with respect to the oxide cladding target MRR cavity (Fig.\ref{fig:results}(a)), but the measurement can still be easily realized with a temperature scan of approximately 10$^\circ C$ . To highlight the importance of the coupling the inset of Fig.\ref{fig:cavity-enhancement}(d) shows the cavity transmission as a function of coupling coefficient $r$. The mismatch between losses and coupling is the discriminator mechanism responsible for the contrast change in response to the change in the losses. We also note that our simulation does not account for a frequency-dependent coupling coefficient , which can compensated by using pulley coupling \cite{moille2019broadband}. Consequently, broadening of the resonances presented in Fig.\ref{fig:cavity-enhancement}(d) and change in the contrast are a direct consequence of the losses. 

\section{Discussion}

Our demonstration marks a significant advancement towards both lab-based and out-of-the lab applications. However, in our current setup, the acquisition and characterization of broadband optical signals was obtained by using of a table-top, large, and non-integrated OSA, hindering the realization of a fully integrated sensor. While chip-scale spectrometers are a prominent filed of research, the coherence of our microcomb suggests the adoption of a dual-comb approach. This method facilitates the conversion of the spectrum and spectroscopic information from the optical to radio-frequency domain, enabling rapid measurement of the entire spectrum. Dual-comb setups with microcombs have been successfully implemented \cite{coddington2016dual}, offering a scalable and integrable solution for broadband detection. Alternatively, a viable, albeit more complex solution, involves using wavelength-division-multiplexer (WDM) filters combined with an array of photodetectors to split the comb signal into separate lines and track them simultaneously.

In our experiment, we utilize identically designed THz-rate MRRs for both comb generation and cavity analysis, facilitating broadband microcomb-MRR coupling with a modest thermo-optic scan range. However, certain sensing scenarios or photonic dispersion characterizations may necessitate finer resolution. Therefore, a more general approach for MRRs of larger radii would be advantageous. Lower FSRs and repetition rates at the GHz rate could potentially enable continuous and simultaneous scanning of our target MRR and microcomb, allowing for gap-less scanning of the entire microcomb spectrum. Achieving this would require establishing the capability to scan the microcomb's frequency offset whilst maintaining a known measurable repetition rate. Additionally, it would be intriguing to explore the possibility of using two MRRs with different radii. In such a scenario, for example, we can leverage the established extreme broadband nature of THz-rate microcombs to measure the dispersion of microcavities with much smaller FSRs (tens and hundreds of GHz), up to the possibility of directly observing zero $D_{int}$ points responsible for dispersion wave generation. Even a relatively sparse THz microcomb frequency grid should be enough to reliably extract smoothly varying dispersion of a GHz-rate microcavities.

Finally, we note that our demonstrated experiment implemented with two separate photonic chips can be straightforwardly integrated onto a single chip, offering a fully integrated solution for microcomb-based cavity-enhanced spectroscopy. Such an integration can also provide a wider characterization bandwidth, eliminating the influence of fiber chip-to-chip interconnects.

\section{Conclusion}
In conclusion, we have proposed and demonstrated the universal photonically-integrated platform based on the microcomb-microcavity coupling, which enables simultaneous investigation of microcavity dispersion and on-chip cavity-enhanced direct frequency comb spectroscopy of molecules. Both comb source and target cavity were implemented using identically designed SiN MRRs. We performed a detailed MRR dispersion characterization by a single-shot temperature scan that outperformed conventional tunable laser source approaches providing a wide bandwidth of ~250 nm. Numerical simulations as well as  measurement by aid of a wavelength meter (albeit, smaller bandwidth) confirmed the reliability of the proposed method. Our measurements provided access to information not only about the cavity resonance positions but also about their lineshapes. Thus, we termed the resulting data representation as complex integrated dispersion. Combination of the comb source and a high-Q cavity along with the complex integrated dispersion representation naturally suggests to use such a system for a cavity-enhanced evanescent spectroscopy. We performed a numerical simulation to demonstrate the expected enhancement of our system, which we evaluated to be at least few dozens in case of a severe losses of the order of 10$\%$  and up to $10^4$ for small losses of the order of 0.01$\%$. Finally, we discussed the apparent limitations of the proposed system and strategies to overcome them. Our study provides a valuable basis for the design and investigation of the MRRs for nonlinear optics applications like soliton or OPO generation, as well as offers a clear path for the implementation of an ultra-compact chip-scale cavity-enhanced spectrometer for biomedical, food-safety \cite{sun2022raman} and environmental sensing applications.

    
\bibliography{sample}

\end{document}